%% file: Lattice2019_86_LEHMANN.tex
\documentclass{PoS}
\usepackage{siunitx}
\usepackage{physics}
\usepackage{gnuplottex}
\usepackage{graphicx}
\usepackage{epstopdf}
\usepackage{sidecap}
\usepackage{bbold}
\newcommand{\iu}{{i\mkern1mu}}
\title{Real-Time-Evolution of Heavy Quarks in the Glasma}

\ShortTitle{Real-Time-Evolution of Heavy Quarks}
\author{\speaker{Alexander Lehmann}$^{\;a,b}$
  {} and Alexander Rothkopf$^{\;b}$\\
  \llap{$^a$} Ruprecht-Karls-Universit\"at Heidelberg, Institut f\"ur Theoretische Physik\\
  Philosophenweg 16, 69120 Heidelberg, Germany\\
  \llap{$^b$} Universitetet i Stavanger, Institutt for matematikk og fysikk\\
  Kristine Bonnevies vei 22, 4021 Stavanger, Norway\\
  E-mail: \email{alexander.lehmann@uis.no}, \email{alexander.rothkopf@uis.no}}

\abstract{We introduce a novel real-time formulation of lattice NRQCD designed for simulations in the background of an highly occupied gluon field. By evolving quarks in the background of a dynamically evolving gluon field we computed the time-evolution of heavy-quarkonium spectral functions as well as the static and for finitely heavy quarks generalised potential. We conclude that the back reaction of the quarks is necessary for any binding process. Here we discuss the methodology, our results and the origin of the absence of a binding process.}

\FullConference{37th International Symposium on Lattice Field Theory - Lattice2019\\
  16-22 June 2019\\
  Wuhan, PR China}

\begin{document}
%%%%%%%%%%%%%%%%%%%%%%%%%%%%%%%%%%%%%%%%%%%%%%
\section{Introduction}
Heavy quarkonia, the bound states of heavy quarks and antiquarks ($b\bar b$ bottomonium, $c\bar c$ charmonium), are one of the main probes for the QGP in heavy ion collisions \cite{Aarts2017}. So far, the formation process of heavy quarkonia in the early time evolution of the glasma remains largely unexplored. The problem lies in the non-perturbative nature of the binding process combined with the necessity to treat the real-time evolution of the system -- which is inaccessible in conventional lattice simulations due to the sign problem.

In this study we employ a hamiltonian evolution scheme in axial gauge with initial conditions drawn from a statistical ensemble for the time-evolution of the gluon field. This approach is called the \emph{classical statistical evolution scheme}, which is formulated in spatial links and electric fields in a non-expanding box. This gauge field is then used as a background for the evolution of heavy quarks and the subsequent evaluation of current-current-correlators to extract heavy quarkonium spectra in the early time evolution of the glasma.
\subsection*{Physical Situation}
The heavy quarks in the glasma are characterized by an hierarchy of energy scales. The relevant scales for the heavy quarks are ordered in powers of the relative velocity $v<\num{1}$. In descending order, they are given by: The heavy quark mass $M$, the heavy quark-antiquark relative momentum $p\sim Mv$ and the corresponding kinetic energy $E\sim Mv^2$. In table \ref{tab:scales} these scales are shown.
\begin{table}[ht]
  \centering
  \begin{tabular}{c|c|c|c|c}
    State & $v^2$ \cite{PhysRevD.89.074011} & $M$ [\SI{}{\giga\electronvolt}] \cite{PhysRevD.98.030001}& $Mv$ [\SI{}{\GeV}]& $Mv^2$ [\SI{}{\GeV}]\\\hline
    $\Upsilon$ & \num{0.1} & \num{4.2} & \num{1.3} & \num{0.42}\\
    $\Psi$ & \num{0.3} & \num{1.3} & \num{0.71} & \num{0.39}
  \end{tabular}
  \caption{Hierarchy of scales for heavy quarkonia of bottom and charm quarks}
  \label{tab:scales}
\end{table}

The initial state of the system is the glasma. Its gluon spectrum can be described \cite{Krasnitz:2001qu} as highly occupied $n(\omega)\sim 1/g^2$ ($g\ll1$) up to the saturation scale $Q_{\text{S}}\approx\SI{1}{\GeV}$. Past that scale, only comparatively small fluctuations from the quantum-1/2 are present.
\subsection*{Method to Study the Far-From-Equilibrium Dynamics}
We are interested in heavy quarkonium spectral functions in the glasma whose form might be time-dependent due to the far-from-equilibrium evolution of the system. In order to take this behaviour into account, they have to be defined as the Wigner distribution function of the corresponding quarkonium current-current-correlator
\begin{align}\label{eq:definition:spectrum}
  \rho\left(t,\omega,\vec{p}=\vec{0}\right)&=\int\limits_{-\infty}^{+\infty}\ev{\comm{J_\mu^a\left(\vec{x},t+\frac{s}{2}\right)}{J^{\mu,a}\left(\vec{x},t-\frac{s}{2}\right)^\dagger}}\text{e}^{-\iu\omega s}\dd{s},
\end{align}
where $J_\mu^a$  is the channel-dependent vector current ($J_\mu^a=(\chi^\dagger M \psi)_\mu^a$ with $M=\mathbb{1}$ for colour and spin singlet, $M=\sigma_i$ for spin triplet etc.), $t$ the central time, $s$ the relative time, $\omega$ the energy of the state and $\vec{p}=\vec{0}$ the relative momentum of the heavy quarks.

Recent progress in the investigation of the early-time gluon-dynamics in heavy ion collisions has been made with \emph{classical statistical} simulations \cite{PhysRevD.87.014026,PhysRevD.98.014006}.
The high occupancy that characterises the glasma makes possible the use of an approximation to the full path integral. It reduces the simulation to a deterministic hamiltonian real-time evolution of gauge fields whose initial conditions are drawn from an ensemble such that they resemble an initial density matrix $\rho_0$. That density matrix is in case of the glasma represented by a box-like occupation number $n(\omega)$. The gauge field $A_\mu^a$ is represented as links $U_\mu(x)=\exp(\iu a_\mu A_\mu^a(x+\hat\mu/2)T^a)$ with $T^a$ the generators of SU(3). The corresponding hamiltonian evolution equations are 
\begin{align}
  U_j\left(t+\frac{a_0}{2},\vec{x}\right)&=\exp\left(\iu a_0a_jE_j^a\left(t,\vec{x}\right)T^a\right)\cdot U_j\left(t-\frac{a_0}{2},\vec{x}\right),\\
  E_j^a\left(t+a_0,\vec{x}\right)&=E_j^a(t,\vec{x})-2\frac{a_0}{a_j}\text{Im}\left\{T^a\sum\limits_{j\neq k}\frac{1}{a_k^2}\left[P_{j,k}\left(t+\frac{a_0}{2},\vec{x}\right)+P_{j,-k}\left(t+\frac{a_0}{2},\vec{x}\right)\right]\right\},
\end{align}
where the plaquettes $P_{j,k}$ and $P_{j,-k}$ are given by
\begin{align}
  P_{j,k}\left(t,\vec{x}\right)&=U_{j}\left(t,\vec{x}\right)U_{k}\left(t,\vec{x}+\hat{j}\right)U_{j}^\dagger\left(t,\vec{x}+\hat{k}\right)U_k^\dagger\left(t,\vec{x}\right),\\
  P_{j,-k}\left(t,\vec{x}\right)&=U_j\left(t,\vec{x}\right)U_k^\dagger\left(t,\vec{x}-\hat{k}+\hat{j}\right)U_j^\dagger\left(t,\vec{x}-\hat{k}\right)U_k\left(t,\vec{x}-\hat{k}\right).
\end{align}

The heavy quarks are described by NRQCD \cite{PhysRevD.43.196,Berwein:2018fos}, an effective field theory that employs the aforementioned hierarchy of scales. It removes the rest mass from the energy scale in which the spectrum is measured and thus allows for much coarser lattice spacings than the prohibitively fine ones necessary to resolve e.g.~the bottomonium production threshold of $2M_{\text{Q}}\approx 8.4\,\text{GeV}$. From the NRQCD-lagrangian follows a non-relativistic Schr\"odinger-equation for the quark (anti-quark) two-point function $G_\psi$ ($G_\chi$)
\begin{equation}\label{eq:twopointfunction_schroedingerequation}
  \pdv{G[U]\left(x_2,x_1\right)}{x_2^0}=-\iu\hat{H}[U](x_2^0)G[U]\left(x_2,x_1\right).
\end{equation}
Here, $U$ denotes the full gauge configuration evolving in Minkowski space-time. On the other hand, $H[U](x_2^0)$ denotes the hamiltonian applied to the two-point function at time $x_2^0$.
We remark that the propagator posesses in addition to its space-time indices denoted in eq.~(\ref{eq:twopointfunction_schroedingerequation}) an additional spin- and colour-structure, i.e.~we are working on the combined hilbert space $\mathcal{H}=\mathcal{H}_{\text{space}}\otimes\mathcal{H}_{\text{spin}}\otimes\mathcal{H}_{\text{colour}}$.
The lattice hamiltonian acting on the two-point-functions G is derived from its continuum expression
\begin{equation}\label{eq:time_dependent_schroedinger_equation}
  \hat{H}_\psi=-\frac{\vec{D}^2}{2m}-\frac{c_1}{2m}B_i\sigma_i-\iu\frac{c_2}{8m^2}\epsilon_{ijk}\{D_i,E_j\}\sigma_k-\frac{c_3}{8m^2}\comm{D_i}{E_i}+\mathcal{O}\left(m^{-3}\right).
\end{equation}
The hamiltonian for the anti-quark field $\chi$ is given by $\hat{H}_\chi(m)=\hat{H}_\psi(-m)$. The various terms in the hamiltonian are formulated such that the corresponding lattice version of the NRQCD-lagrangian is gauge-invariant.
The time-dependent Schr\"odinger-equation in eq.~(\ref{eq:time_dependent_schroedinger_equation}) is numerically intergrated with the Crank-Nicolson method. The links and their corresponding momenta, the chromo-electric field $E_i^a$, are evolved with a Leapfrog method, indicated by the shift of $a_0/2$ in the link variables.

Regarding the interplay of gluons and the heavy quarks, the much bigger heavy quark-antiquark production threshold $2M$ has to be compared to the saturation scale $Q_{\text{S}}$. In the initial state, due to the lack of gluon occupation at $\omega\sim 2M$, pair production of heavy quarks from gluons is a strongly suppressed process and thus allows to remove the vacuum polarisation effects from the simulation by setting the heavy-quark fermion determinant to 1.

%%%%%%%%%%%%%%%%%%%%%%%%%%%%%%%%%%%%%%%%%%%%%%
\section{Extraction of the Real-Time Spectrum}
In eq.~(\ref{eq:definition:spectrum}) the formal definition of the spectral function via the commutator of the vector-current was given. Expanding the commutator leaves us with the two current-current correlators in the evolving glasma $\ev{JJ^\dagger}_{\rho_{\text{Glasma}}}$ and $\ev{J^\dagger J}_{\rho_{\text{Glasma}}}$. We argue that the second expectation value can be neglected due to the suppression of pair creation for heavy quark-antiquark-pairs due to the gluonic background field that has an occupation up to $Q_{\text{S}}\approx\SI{1.2}{\GeV}$ only. Thus, we are left with the computation of $\ev{JJ^\dagger}_{\rho_{\text{Glasma}}}$.
After inserting the interpolators from \cite{PhysRevD.43.196} and integrating out the heavy quark fields one obtains the formulas for the quarkonium correlators.
%%%%%%%%%%%%%%%%%%%%%%%%%%%%%%%%%%%%%%%%%%%%%%
\section{Conceptual Differences to the Euclidean Lattice Method}
There are several conceptual differences to the Euclidean lattice field theory that show up in the correlator and its analysis which we will discuss in this section.

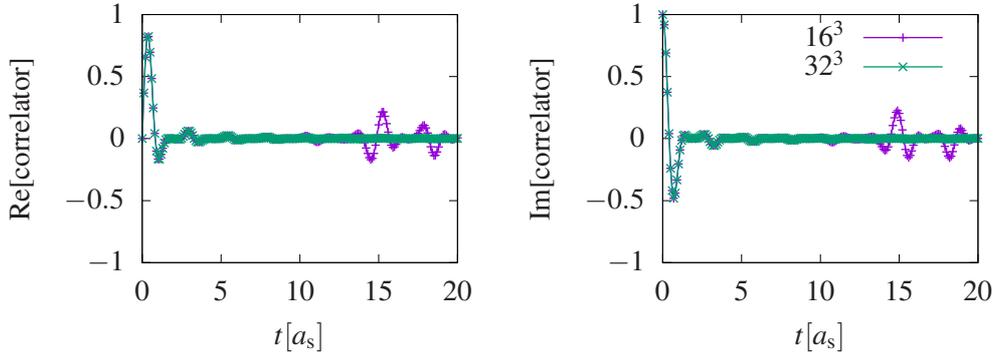
\begin{figure}
  \centering
      {\graphicspath{{./plots/}}\input{plots/plot1.tex}}
      {\graphicspath{{./plots/}}\input{plots/plot2.tex}}
      \caption{Real (left) and imaginary (right) part of the free heavy quarkonium correlator in real time for $a_{\text{s}}M_{\text{Q}}=\pi/2$, ${a_0/a_{\text{s}}=0.1}$ and $\Lambda\in\{16^3,32^3\}$}
      \label{fig:freecorrelator}
\end{figure}

\textbf{Real-time signals} show an oscillatory behaviour that arises from the structure of the Schr\"odinger-equation. In fig.~\ref{fig:freecorrelator} the real and imaginary part of the free, i.e.~${U_\mu\equiv\mathbb{1}}$ and ${E_i\equiv0}$, heavy quarkonium correlator are shown for a temporal spacing of $a_0/a_s=0.1$, the two lattice sizes $\Lambda=16^3$ and $\Lambda=32^3$ and the quark mass $a_{\text{s}}M_{\text{Q}}=\pi/2$. In general, the real-time correlator is a-priori infinite and not related to a temperature.

\textbf{Finite volume effects} arise as recurrence in the correlator. This can be understood by considering the diffusive term $-D^2/(2m)$ in the hamiltonian for the two-point-correlator $G$ that leads to a decay in time on top of any oscillatory behaviour. At a finite volume, however, an initial distributions diffusion surpasses the boundary and causes recurrence. This recurrence is the real-time finite-volume effect and leads to an additional peak structure in the spectrum when evolving to late times and thus resolving the seperate, discrete lattice momenta. In fig.~\ref{fig:freecorrelator} the decay can be clearly seen on top of the aforementioned oscillations. At circa $t=10a_s$ the real and imaginary parts of the two correlator on the $16^3$ and $32^3$ lattices start to deviate from each other due to the arising recurrence on the smaller lattice.

\textbf{The extraction of the spectrum} has to be done via a Fourier transform as follows from eq.~(\ref{eq:definition:spectrum}) instead of a Laplace transform. While the inverse Laplace-transform for noisy data is an ill-posed problem, the inverse Fourier transform does not suffer from this \cite{Craig:1994}.

\textbf{The removal of finite time artifacts from the spectrum} can be done via \emph{windowing}. The real-time signal would a-priori have to be sampled over the whole real axis - something, we cannot do. In praxis we have to stop the measurement at some point and thus cut the signals extension to the window $t\in[t_{\text{i}},t_{\text{f}}]$. This amounts to multiplying the signal with the window-function $f(t_{\text{i}},t_{\text{f}};t)=\Theta(t_{\text{i}}-t)\Theta(t_{\text{f}}-t)$. The subsequent Fourier transform leaves us with the convolution of the signals and the window functions seperate Fourier transforms which leads to so-called \emph{ringing}. This effect can be partially compensated by choosing instead a window-function that reduces the effect of the hard cut-offs at $t_{\text{i}}$ and $t_{\text{f}}$. The window-function we use is called the Hann-window. It smoothly sets the signal to 0 at the cut-offs while retaining the signals form during most of the interval over which the signal has been measured.
%%%%%%%%%%%%%%%%%%%%%%%%%%%%%%%%%%%%%%%%%%%%%%
\section{Numerical Results}
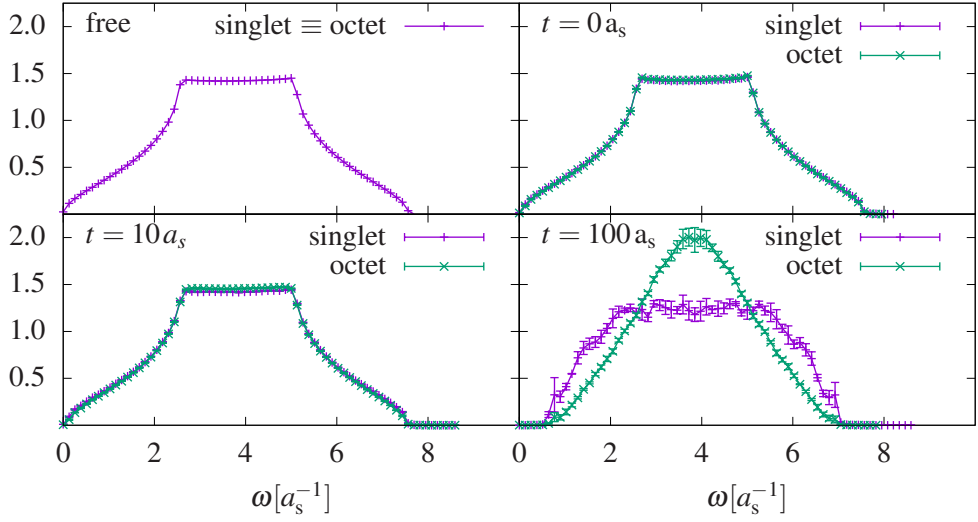
\begin{figure}
  \centering
      {\graphicspath{{./plots/}}\input{plots/plot3.tex}}
      \caption{Colour singlet (violet ticks) and octet (green crosses) heavy quarkonium spectrum for $a_{\text{s}}Q_{\text{s}}=1$, $a_{\text{s}}M_{\text{Q}}=\pi/2$, $a_0/a_{\text{s}}=0.1$ on a $64^3$-lattice. From $Q_{\text{S}}/M_Q=1.2$ follows that we consider here slightly heavier charm quarks.}
      \label{fig:spectra}
\end{figure}

We start with the free spectrum. In the top left plot of fig.~\ref{fig:spectra} the free S-wave spectrum for an heavy quarkonium is shown. The scale is set by the gluon saturation scale $Q_{\text{s}}$: Via $a_{\text{s}}=(a_{\text{s}}/Q_{\text{s}})\cdot Q_{\text{s}}$ we can determine the physical values of all other quantities. We used here quark masses of $M_{\text{Q}}=\pi/2\cdot Q_{\text{s}}\approx\SI{1.9}{\GeV}$ and a lattice spacing of $a_{\text{s}}=\SI{0.8}{\per\GeV}$. The specific form of the free spectrum originates in the use of the Crank-Nicolson-scheme and the value of the quark mass in units of the lattice spacing. The \emph{cusps} of the spectral function are finite-lattice-spacing- and thus momentum-cut-off-artifacts. Therefore, we can only draw physical implications from the spectral functions until at most the first cusp which is in the shown plots at around $\omega=\SI{2.5}{a_{\text{s}}^{-1}}$.

In the same figure, we show the colour singlet and octet heavy quarkonium spectral functions at $t=0,10,100$ spatial lattice spacings. We can see that the free, singlet and octet spectral functions agree for $t=\SI{0}{a_{\text{s}}}$. At $t=\SI{10}{a_{\text{s}}}$ we start to see a slight deviation of the singlet and octet spectral functions. At the latest shown time, $t=\SI{100}{a_{\text{s}}}$, we see a clear difference between the two spectral functions and from the free one. In the energy-interval $a_{\text{s}}\omega\in[0,2]$ we see an enhancement of the singlet compared to the octet channel but no peak structure. We understand the absence of for bound states characteristic peaks from the missing back reaction of the quarks to the gluon field: In contrast to the Euclidean case, where this back reaction happens solely through the fermion determinant, in a classical statistical simulation the fermions appear also in the Maxwell equations and the Gauss law constraint which reduces the allowed configuration space. Therefore, even in a real-time simulation of a gluon field with static quarks, the projection of the physical space to one that fulfils the Gauss constraint would have to be performed in order to have the chance to see any kind of binding process. 

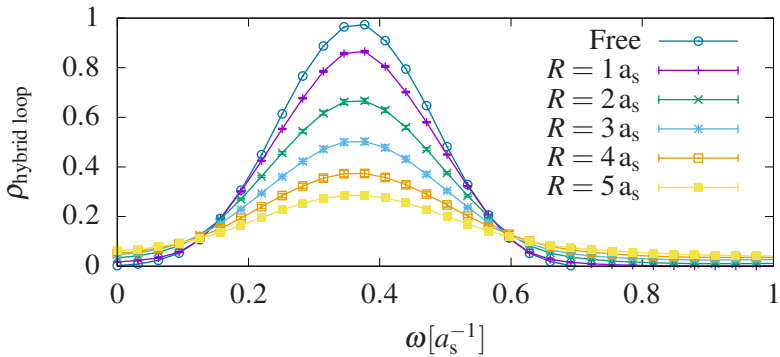
\begin{SCfigure}
  \centering
      {\graphicspath{{./plots/}}\input{plots/plot4.tex}}
      \caption{Hybrid loops spectral function on a $12^3$-lattice for $\beta=16$ and $a_{\text{s}}M_{\text{Q}}=10\pi/2$}
      \label{fig:hybridloopspectrum}
\end{SCfigure}

In order to support that argument, we computed the static quark potential from the Wilson loop as it was done in \cite{Laine:2006ns}, where the initial states were drawn from a classical thermal ensemble. The static quarkonium potential was found to lack the presence of a real part, thus indicating the absence of binding. In order to investigate wether the origin of this absence lies in the static nature of the quark fields, we considered quarks with a finite but still heavy mass using our evolution scheme. Therefore, we replaced the temporal Wilson lines by the quark and anti-quark propagators and thus computed a generalised, hybrid Wilson loop. By investigating its spectrum, as it was done for the conventional Wilson loop in \cite{Burnier:2012az}, we can extract the potential from the peaks positions and its imaginary part from the peaks widths. In fig.~\ref{fig:hybridloopspectrum} the hybrid loops spectrum for several distances are shown for the same parameters as in \cite{Laine:2006ns}, i.e.~on a $12^3$-lattice for $\beta=16$, and additionally with $a_{\text{s}}M_{\text{Q}}=10\pi/2$. We see that the peaks are located at an energy different from zero but do not change position with spatial separation of the loop. However, because the quarks have a finite mass, they can move and thus the energy receives a contribution from a finite kinetic part, ${E=p^2/2M_{\text{Q}} + \text{Re}V(R)}$, which for the free case reduces to ${E_{\text{free}}=p^2/2M_{\text{Q}}}$. Hence the shift in the spectral features can be understood as an irrelevant constant contribution due to the presence of the kinetic term. We conclude that the real part of the potential even for finite mass remains zero in simulations without back reaction.

%%%%%%%%%%%%%%%%%%%%%%%%%%%%%%%%%%%%%%%%%%%%%%
\section{Conclusion and Outlook}
We constructed a novel simulation method for heavy quarks in the background of a dynamically evolving gluon field. First, we investigated the spectrum of heavy quarkonium current-current-correlators and found no indications of bound states. We explained this by the absence of the back reaction from the quarks to the gluon field. We supporteded this argument by the lack of a real part in the static as well as in the for heavy quarks generalised potential.

In the future we aim at the inclusion of the back reaction from the quarks to the gluon field. As a first step, we are currently working on the computation of the static potential in the presence of a dynamically evolving gauge field.

%%%%%%%%%%%%%%
\section*{Acknowledgments}
We thank Y.~Akamatsu, K.~Boguslavski, H.~Fujii and M.~Kitazawa for insightful discussions. A.~Lehmann thanks the Nuclear Theory Group at Osaka University for their hospitality. This work is supported by the DFG Collaborative Research Centre "SFB 1225 (ISOQUANT)". The authors acknowledge a generous allocation of computing resources by UNINETT Sigma2 AS, Norway.
\bibliographystyle{JHEP}
\bibliography{Lattice2019_86_LEHMANN}
\end{document}

%% file: plots/plot1.tex
% GNUPLOT: LaTeX picture with Postscript
\begingroup
  \makeatletter
  \providecommand\color[2][]{%
    \GenericError{(gnuplot) \space\space\space\@spaces}{%
      Package color not loaded in conjunction with
      terminal option `colourtext'%
    }{See the gnuplot documentation for explanation.%
    }{Either use 'blacktext' in gnuplot or load the package
      color.sty in LaTeX.}%
    \renewcommand\color[2][]{}%
  }%
  \providecommand\includegraphics[2][]{%
    \GenericError{(gnuplot) \space\space\space\@spaces}{%
      Package graphicx or graphics not loaded%
    }{See the gnuplot documentation for explanation.%
    }{The gnuplot epslatex terminal needs graphicx.sty or graphics.sty.}%
    \renewcommand\includegraphics[2][]{}%
  }%
  \providecommand\rotatebox[2]{#2}%
  \@ifundefined{ifGPcolor}{%
    \newif\ifGPcolor
    \GPcolortrue
  }{}%
  \@ifundefined{ifGPblacktext}{%
    \newif\ifGPblacktext
    \GPblacktexttrue
  }{}%
  % define a \g@addto@macro without @ in the name:
  \let\gplgaddtomacro\g@addto@macro
  % define empty templates for all commands taking text:
  \gdef\gplbacktext{}%
  \gdef\gplfronttext{}%
  \makeatother
  \ifGPblacktext
    % no textcolor at all
    \def\colorrgb#1{}%
    \def\colorgray#1{}%
  \else
    % gray or color?
    \ifGPcolor
      \def\colorrgb#1{\color[rgb]{#1}}%
      \def\colorgray#1{\color[gray]{#1}}%
      \expandafter\def\csname LTw\endcsname{\color{white}}%
      \expandafter\def\csname LTb\endcsname{\color{black}}%
      \expandafter\def\csname LTa\endcsname{\color{black}}%
      \expandafter\def\csname LT0\endcsname{\color[rgb]{1,0,0}}%
      \expandafter\def\csname LT1\endcsname{\color[rgb]{0,1,0}}%
      \expandafter\def\csname LT2\endcsname{\color[rgb]{0,0,1}}%
      \expandafter\def\csname LT3\endcsname{\color[rgb]{1,0,1}}%
      \expandafter\def\csname LT4\endcsname{\color[rgb]{0,1,1}}%
      \expandafter\def\csname LT5\endcsname{\color[rgb]{1,1,0}}%
      \expandafter\def\csname LT6\endcsname{\color[rgb]{0,0,0}}%
      \expandafter\def\csname LT7\endcsname{\color[rgb]{1,0.3,0}}%
      \expandafter\def\csname LT8\endcsname{\color[rgb]{0.5,0.5,0.5}}%
    \else
      % gray
      \def\colorrgb#1{\color{black}}%
      \def\colorgray#1{\color[gray]{#1}}%
      \expandafter\def\csname LTw\endcsname{\color{white}}%
      \expandafter\def\csname LTb\endcsname{\color{black}}%
      \expandafter\def\csname LTa\endcsname{\color{black}}%
      \expandafter\def\csname LT0\endcsname{\color{black}}%
      \expandafter\def\csname LT1\endcsname{\color{black}}%
      \expandafter\def\csname LT2\endcsname{\color{black}}%
      \expandafter\def\csname LT3\endcsname{\color{black}}%
      \expandafter\def\csname LT4\endcsname{\color{black}}%
      \expandafter\def\csname LT5\endcsname{\color{black}}%
      \expandafter\def\csname LT6\endcsname{\color{black}}%
      \expandafter\def\csname LT7\endcsname{\color{black}}%
      \expandafter\def\csname LT8\endcsname{\color{black}}%
    \fi
  \fi
    \setlength{\unitlength}{0.0500bp}%
    \ifx\gptboxheight\undefined%
      \newlength{\gptboxheight}%
      \newlength{\gptboxwidth}%
      \newsavebox{\gptboxtext}%
    \fi%
    \setlength{\fboxrule}{0.5pt}%
    \setlength{\fboxsep}{1pt}%
\begin{picture}(3826.00,2834.00)%
    \gplgaddtomacro\gplbacktext{%
      \csname LTb\endcsname%
      \put(946,704){\makebox(0,0)[r]{\strut{}$-1$}}%
      \put(946,1170){\makebox(0,0)[r]{\strut{}$-0.5$}}%
      \put(946,1637){\makebox(0,0)[r]{\strut{}$0$}}%
      \put(946,2103){\makebox(0,0)[r]{\strut{}$0.5$}}%
      \put(946,2569){\makebox(0,0)[r]{\strut{}$1$}}%
      \put(1078,484){\makebox(0,0){\strut{}$0$}}%
      \put(1666,484){\makebox(0,0){\strut{}$5$}}%
      \put(2254,484){\makebox(0,0){\strut{}$10$}}%
      \put(2841,484){\makebox(0,0){\strut{}$15$}}%
      \put(3429,484){\makebox(0,0){\strut{}$20$}}%
    }%
    \gplgaddtomacro\gplfronttext{%
      \csname LTb\endcsname%
      \put(176,1636){\rotatebox{-270}{\makebox(0,0){\strut{}Re[correlator]}}}%
      \put(2253,154){\makebox(0,0){\strut{}$t [a_{\text{s}}]$}}%
    }%
    \gplbacktext
    \put(0,0){\includegraphics{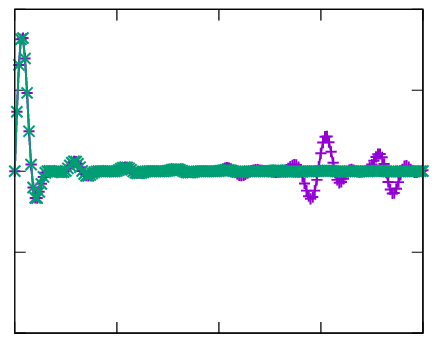}}%
    \gplfronttext
  \end{picture}%
\endgroup

%% file: plots/plot2.tex
% GNUPLOT: LaTeX picture with Postscript
\begingroup
  \makeatletter
  \providecommand\color[2][]{%
    \GenericError{(gnuplot) \space\space\space\@spaces}{%
      Package color not loaded in conjunction with
      terminal option `colourtext'%
    }{See the gnuplot documentation for explanation.%
    }{Either use 'blacktext' in gnuplot or load the package
      color.sty in LaTeX.}%
    \renewcommand\color[2][]{}%
  }%
  \providecommand\includegraphics[2][]{%
    \GenericError{(gnuplot) \space\space\space\@spaces}{%
      Package graphicx or graphics not loaded%
    }{See the gnuplot documentation for explanation.%
    }{The gnuplot epslatex terminal needs graphicx.sty or graphics.sty.}%
    \renewcommand\includegraphics[2][]{}%
  }%
  \providecommand\rotatebox[2]{#2}%
  \@ifundefined{ifGPcolor}{%
    \newif\ifGPcolor
    \GPcolortrue
  }{}%
  \@ifundefined{ifGPblacktext}{%
    \newif\ifGPblacktext
    \GPblacktexttrue
  }{}%
  % define a \g@addto@macro without @ in the name:
  \let\gplgaddtomacro\g@addto@macro
  % define empty templates for all commands taking text:
  \gdef\gplbacktext{}%
  \gdef\gplfronttext{}%
  \makeatother
  \ifGPblacktext
    % no textcolor at all
    \def\colorrgb#1{}%
    \def\colorgray#1{}%
  \else
    % gray or color?
    \ifGPcolor
      \def\colorrgb#1{\color[rgb]{#1}}%
      \def\colorgray#1{\color[gray]{#1}}%
      \expandafter\def\csname LTw\endcsname{\color{white}}%
      \expandafter\def\csname LTb\endcsname{\color{black}}%
      \expandafter\def\csname LTa\endcsname{\color{black}}%
      \expandafter\def\csname LT0\endcsname{\color[rgb]{1,0,0}}%
      \expandafter\def\csname LT1\endcsname{\color[rgb]{0,1,0}}%
      \expandafter\def\csname LT2\endcsname{\color[rgb]{0,0,1}}%
      \expandafter\def\csname LT3\endcsname{\color[rgb]{1,0,1}}%
      \expandafter\def\csname LT4\endcsname{\color[rgb]{0,1,1}}%
      \expandafter\def\csname LT5\endcsname{\color[rgb]{1,1,0}}%
      \expandafter\def\csname LT6\endcsname{\color[rgb]{0,0,0}}%
      \expandafter\def\csname LT7\endcsname{\color[rgb]{1,0.3,0}}%
      \expandafter\def\csname LT8\endcsname{\color[rgb]{0.5,0.5,0.5}}%
    \else
      % gray
      \def\colorrgb#1{\color{black}}%
      \def\colorgray#1{\color[gray]{#1}}%
      \expandafter\def\csname LTw\endcsname{\color{white}}%
      \expandafter\def\csname LTb\endcsname{\color{black}}%
      \expandafter\def\csname LTa\endcsname{\color{black}}%
      \expandafter\def\csname LT0\endcsname{\color{black}}%
      \expandafter\def\csname LT1\endcsname{\color{black}}%
      \expandafter\def\csname LT2\endcsname{\color{black}}%
      \expandafter\def\csname LT3\endcsname{\color{black}}%
      \expandafter\def\csname LT4\endcsname{\color{black}}%
      \expandafter\def\csname LT5\endcsname{\color{black}}%
      \expandafter\def\csname LT6\endcsname{\color{black}}%
      \expandafter\def\csname LT7\endcsname{\color{black}}%
      \expandafter\def\csname LT8\endcsname{\color{black}}%
    \fi
  \fi
    \setlength{\unitlength}{0.0500bp}%
    \ifx\gptboxheight\undefined%
      \newlength{\gptboxheight}%
      \newlength{\gptboxwidth}%
      \newsavebox{\gptboxtext}%
    \fi%
    \setlength{\fboxrule}{0.5pt}%
    \setlength{\fboxsep}{1pt}%
\begin{picture}(3826.00,2834.00)%
    \gplgaddtomacro\gplbacktext{%
      \csname LTb\endcsname%
      \put(946,704){\makebox(0,0)[r]{\strut{}$-1$}}%
      \put(946,1170){\makebox(0,0)[r]{\strut{}$-0.5$}}%
      \put(946,1637){\makebox(0,0)[r]{\strut{}$0$}}%
      \put(946,2103){\makebox(0,0)[r]{\strut{}$0.5$}}%
      \put(946,2569){\makebox(0,0)[r]{\strut{}$1$}}%
      \put(1078,484){\makebox(0,0){\strut{}$0$}}%
      \put(1666,484){\makebox(0,0){\strut{}$5$}}%
      \put(2254,484){\makebox(0,0){\strut{}$10$}}%
      \put(2841,484){\makebox(0,0){\strut{}$15$}}%
      \put(3429,484){\makebox(0,0){\strut{}$20$}}%
    }%
    \gplgaddtomacro\gplfronttext{%
      \csname LTb\endcsname%
      \put(176,1636){\rotatebox{-270}{\makebox(0,0){\strut{}Im[correlator]}}}%
      \put(2253,154){\makebox(0,0){\strut{}$t [a_{\text{s}}]$}}%
      \csname LTb\endcsname%
      \put(2442,2396){\makebox(0,0)[r]{\strut{}$16^3$}}%
      \csname LTb\endcsname%
      \put(2442,2176){\makebox(0,0)[r]{\strut{}$32^3$}}%
    }%
    \gplbacktext
    \put(0,0){\includegraphics{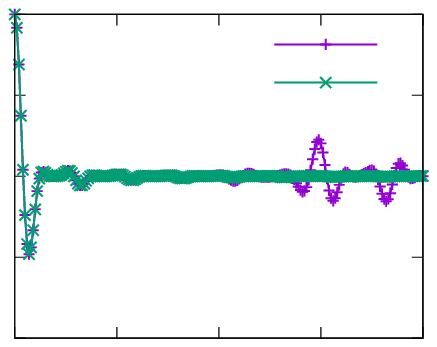}}%
    \gplfronttext
  \end{picture}%
\endgroup

%% file: plots/plot3.tex
% GNUPLOT: LaTeX picture with Postscript
\begingroup
  \makeatletter
  \providecommand\color[2][]{%
    \GenericError{(gnuplot) \space\space\space\@spaces}{%
      Package color not loaded in conjunction with
      terminal option `colourtext'%
    }{See the gnuplot documentation for explanation.%
    }{Either use 'blacktext' in gnuplot or load the package
      color.sty in LaTeX.}%
    \renewcommand\color[2][]{}%
  }%
  \providecommand\includegraphics[2][]{%
    \GenericError{(gnuplot) \space\space\space\@spaces}{%
      Package graphicx or graphics not loaded%
    }{See the gnuplot documentation for explanation.%
    }{The gnuplot epslatex terminal needs graphicx.sty or graphics.sty.}%
    \renewcommand\includegraphics[2][]{}%
  }%
  \providecommand\rotatebox[2]{#2}%
  \@ifundefined{ifGPcolor}{%
    \newif\ifGPcolor
    \GPcolortrue
  }{}%
  \@ifundefined{ifGPblacktext}{%
    \newif\ifGPblacktext
    \GPblacktexttrue
  }{}%
  % define a \g@addto@macro without @ in the name:
  \let\gplgaddtomacro\g@addto@macro
  % define empty templates for all commands taking text:
  \gdef\gplbacktext{}%
  \gdef\gplfronttext{}%
  \makeatother
  \ifGPblacktext
    % no textcolor at all
    \def\colorrgb#1{}%
    \def\colorgray#1{}%
  \else
    % gray or color?
    \ifGPcolor
      \def\colorrgb#1{\color[rgb]{#1}}%
      \def\colorgray#1{\color[gray]{#1}}%
      \expandafter\def\csname LTw\endcsname{\color{white}}%
      \expandafter\def\csname LTb\endcsname{\color{black}}%
      \expandafter\def\csname LTa\endcsname{\color{black}}%
      \expandafter\def\csname LT0\endcsname{\color[rgb]{1,0,0}}%
      \expandafter\def\csname LT1\endcsname{\color[rgb]{0,1,0}}%
      \expandafter\def\csname LT2\endcsname{\color[rgb]{0,0,1}}%
      \expandafter\def\csname LT3\endcsname{\color[rgb]{1,0,1}}%
      \expandafter\def\csname LT4\endcsname{\color[rgb]{0,1,1}}%
      \expandafter\def\csname LT5\endcsname{\color[rgb]{1,1,0}}%
      \expandafter\def\csname LT6\endcsname{\color[rgb]{0,0,0}}%
      \expandafter\def\csname LT7\endcsname{\color[rgb]{1,0.3,0}}%
      \expandafter\def\csname LT8\endcsname{\color[rgb]{0.5,0.5,0.5}}%
    \else
      % gray
      \def\colorrgb#1{\color{black}}%
      \def\colorgray#1{\color[gray]{#1}}%
      \expandafter\def\csname LTw\endcsname{\color{white}}%
      \expandafter\def\csname LTb\endcsname{\color{black}}%
      \expandafter\def\csname LTa\endcsname{\color{black}}%
      \expandafter\def\csname LT0\endcsname{\color{black}}%
      \expandafter\def\csname LT1\endcsname{\color{black}}%
      \expandafter\def\csname LT2\endcsname{\color{black}}%
      \expandafter\def\csname LT3\endcsname{\color{black}}%
      \expandafter\def\csname LT4\endcsname{\color{black}}%
      \expandafter\def\csname LT5\endcsname{\color{black}}%
      \expandafter\def\csname LT6\endcsname{\color{black}}%
      \expandafter\def\csname LT7\endcsname{\color{black}}%
      \expandafter\def\csname LT8\endcsname{\color{black}}%
    \fi
  \fi
    \setlength{\unitlength}{0.0500bp}%
    \ifx\gptboxheight\undefined%
      \newlength{\gptboxheight}%
      \newlength{\gptboxwidth}%
      \newsavebox{\gptboxtext}%
    \fi%
    \setlength{\fboxrule}{0.5pt}%
    \setlength{\fboxsep}{1pt}%
\begin{picture}(8502.00,4534.00)%
    \gplgaddtomacro\gplbacktext{%
      \csname LTb\endcsname%
      \put(1143,2845){\makebox(0,0)[r]{\strut{}0.5}}%
      \put(1143,3198){\makebox(0,0)[r]{\strut{}1.0}}%
      \put(1143,3550){\makebox(0,0)[r]{\strut{}1.5}}%
      \put(1143,3903){\makebox(0,0)[r]{\strut{}2.0}}%
      \put(1275,2273){\makebox(0,0){\strut{}}}%
      \put(1955,2273){\makebox(0,0){\strut{}}}%
      \put(2635,2273){\makebox(0,0){\strut{}}}%
      \put(3315,2273){\makebox(0,0){\strut{}}}%
      \put(3995,2273){\makebox(0,0){\strut{}}}%
      \put(1445,3920){\makebox(0,0)[l]{\strut{}free}}%
    }%
    \gplgaddtomacro\gplfronttext{%
      \csname LTb\endcsname%
      \put(3688,3906){\makebox(0,0)[r]{\strut{}singlet $\equiv$ octet}}%
    }%
    \gplgaddtomacro\gplbacktext{%
      \csname LTb\endcsname%
      \put(4544,2845){\makebox(0,0)[r]{\strut{}}}%
      \put(4544,3198){\makebox(0,0)[r]{\strut{}}}%
      \put(4544,3550){\makebox(0,0)[r]{\strut{}}}%
      \put(4544,3903){\makebox(0,0)[r]{\strut{}}}%
      \put(4676,2273){\makebox(0,0){\strut{}}}%
      \put(5356,2273){\makebox(0,0){\strut{}}}%
      \put(6036,2273){\makebox(0,0){\strut{}}}%
      \put(6715,2273){\makebox(0,0){\strut{}}}%
      \put(7395,2273){\makebox(0,0){\strut{}}}%
      \put(4846,3920){\makebox(0,0)[l]{\strut{}$t=\SI{0}{a_{\text{s}}}$}}%
    }%
    \gplgaddtomacro\gplfronttext{%
      \csname LTb\endcsname%
      \put(7088,3906){\makebox(0,0)[r]{\strut{}singlet}}%
      \csname LTb\endcsname%
      \put(7088,3686){\makebox(0,0)[r]{\strut{}octet}}%
    }%
    \gplgaddtomacro\gplbacktext{%
      \csname LTb\endcsname%
      \put(1143,1259){\makebox(0,0)[r]{\strut{}0.5}}%
      \put(1143,1611){\makebox(0,0)[r]{\strut{}1.0}}%
      \put(1143,1964){\makebox(0,0)[r]{\strut{}1.5}}%
      \put(1143,2317){\makebox(0,0)[r]{\strut{}2.0}}%
      \put(1275,686){\makebox(0,0){\strut{}0}}%
      \put(1955,686){\makebox(0,0){\strut{}2}}%
      \put(2635,686){\makebox(0,0){\strut{}4}}%
      \put(3315,686){\makebox(0,0){\strut{}6}}%
      \put(3995,686){\makebox(0,0){\strut{}8}}%
      \put(1445,2334){\makebox(0,0)[l]{\strut{}$ t=10\,a_s$}}%
    }%
    \gplgaddtomacro\gplfronttext{%
      \csname LTb\endcsname%
      \put(2975,356){\makebox(0,0){\strut{}$\omega [a_{\text{s}}^{-1}]$}}%
      \csname LTb\endcsname%
      \put(3688,2320){\makebox(0,0)[r]{\strut{}singlet}}%
      \csname LTb\endcsname%
      \put(3688,2100){\makebox(0,0)[r]{\strut{}octet}}%
    }%
    \gplgaddtomacro\gplbacktext{%
      \csname LTb\endcsname%
      \put(4544,1259){\makebox(0,0)[r]{\strut{}}}%
      \put(4544,1611){\makebox(0,0)[r]{\strut{}}}%
      \put(4544,1964){\makebox(0,0)[r]{\strut{}}}%
      \put(4544,2317){\makebox(0,0)[r]{\strut{}}}%
      \put(4676,686){\makebox(0,0){\strut{}0}}%
      \put(5356,686){\makebox(0,0){\strut{}2}}%
      \put(6036,686){\makebox(0,0){\strut{}4}}%
      \put(6715,686){\makebox(0,0){\strut{}6}}%
      \put(7395,686){\makebox(0,0){\strut{}8}}%
      \put(4846,2334){\makebox(0,0)[l]{\strut{}$t=\SI{100}{a_{\text{s}}}$}}%
    }%
    \gplgaddtomacro\gplfronttext{%
      \csname LTb\endcsname%
      \put(6375,356){\makebox(0,0){\strut{}$ \omega [a_{\text{s}}^{-1}]$}}%
      \csname LTb\endcsname%
      \put(7088,2320){\makebox(0,0)[r]{\strut{}singlet}}%
      \csname LTb\endcsname%
      \put(7088,2100){\makebox(0,0)[r]{\strut{}octet}}%
    }%
    \gplbacktext
    \put(0,0){\includegraphics{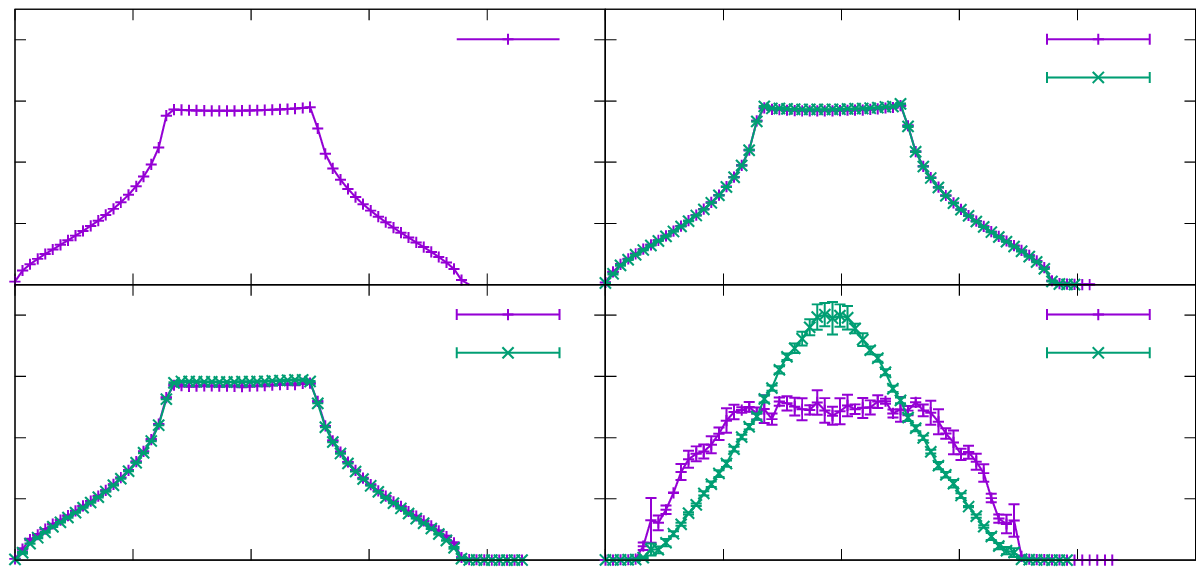}}%
    \gplfronttext
  \end{picture}%
\endgroup

%% file: plots/plot4.tex
% GNUPLOT: LaTeX picture with Postscript
\begingroup
  \makeatletter
  \providecommand\color[2][]{%
    \GenericError{(gnuplot) \space\space\space\@spaces}{%
      Package color not loaded in conjunction with
      terminal option `colourtext'%
    }{See the gnuplot documentation for explanation.%
    }{Either use 'blacktext' in gnuplot or load the package
      color.sty in LaTeX.}%
    \renewcommand\color[2][]{}%
  }%
  \providecommand\includegraphics[2][]{%
    \GenericError{(gnuplot) \space\space\space\@spaces}{%
      Package graphicx or graphics not loaded%
    }{See the gnuplot documentation for explanation.%
    }{The gnuplot epslatex terminal needs graphicx.sty or graphics.sty.}%
    \renewcommand\includegraphics[2][]{}%
  }%
  \providecommand\rotatebox[2]{#2}%
  \@ifundefined{ifGPcolor}{%
    \newif\ifGPcolor
    \GPcolortrue
  }{}%
  \@ifundefined{ifGPblacktext}{%
    \newif\ifGPblacktext
    \GPblacktexttrue
  }{}%
  % define a \g@addto@macro without @ in the name:
  \let\gplgaddtomacro\g@addto@macro
  % define empty templates for all commands taking text:
  \gdef\gplbacktext{}%
  \gdef\gplfronttext{}%
  \makeatother
  \ifGPblacktext
    % no textcolor at all
    \def\colorrgb#1{}%
    \def\colorgray#1{}%
  \else
    % gray or color?
    \ifGPcolor
      \def\colorrgb#1{\color[rgb]{#1}}%
      \def\colorgray#1{\color[gray]{#1}}%
      \expandafter\def\csname LTw\endcsname{\color{white}}%
      \expandafter\def\csname LTb\endcsname{\color{black}}%
      \expandafter\def\csname LTa\endcsname{\color{black}}%
      \expandafter\def\csname LT0\endcsname{\color[rgb]{1,0,0}}%
      \expandafter\def\csname LT1\endcsname{\color[rgb]{0,1,0}}%
      \expandafter\def\csname LT2\endcsname{\color[rgb]{0,0,1}}%
      \expandafter\def\csname LT3\endcsname{\color[rgb]{1,0,1}}%
      \expandafter\def\csname LT4\endcsname{\color[rgb]{0,1,1}}%
      \expandafter\def\csname LT5\endcsname{\color[rgb]{1,1,0}}%
      \expandafter\def\csname LT6\endcsname{\color[rgb]{0,0,0}}%
      \expandafter\def\csname LT7\endcsname{\color[rgb]{1,0.3,0}}%
      \expandafter\def\csname LT8\endcsname{\color[rgb]{0.5,0.5,0.5}}%
    \else
      % gray
      \def\colorrgb#1{\color{black}}%
      \def\colorgray#1{\color[gray]{#1}}%
      \expandafter\def\csname LTw\endcsname{\color{white}}%
      \expandafter\def\csname LTb\endcsname{\color{black}}%
      \expandafter\def\csname LTa\endcsname{\color{black}}%
      \expandafter\def\csname LT0\endcsname{\color{black}}%
      \expandafter\def\csname LT1\endcsname{\color{black}}%
      \expandafter\def\csname LT2\endcsname{\color{black}}%
      \expandafter\def\csname LT3\endcsname{\color{black}}%
      \expandafter\def\csname LT4\endcsname{\color{black}}%
      \expandafter\def\csname LT5\endcsname{\color{black}}%
      \expandafter\def\csname LT6\endcsname{\color{black}}%
      \expandafter\def\csname LT7\endcsname{\color{black}}%
      \expandafter\def\csname LT8\endcsname{\color{black}}%
    \fi
  \fi
    \setlength{\unitlength}{0.0500bp}%
    \ifx\gptboxheight\undefined%
      \newlength{\gptboxheight}%
      \newlength{\gptboxwidth}%
      \newsavebox{\gptboxtext}%
    \fi%
    \setlength{\fboxrule}{0.5pt}%
    \setlength{\fboxsep}{1pt}%
\begin{picture}(6236.00,2834.00)%
    \gplgaddtomacro\gplbacktext{%
      \csname LTb\endcsname%
      \put(814,704){\makebox(0,0)[r]{\strut{}$0$}}%
      \put(814,1077){\makebox(0,0)[r]{\strut{}$0.2$}}%
      \put(814,1450){\makebox(0,0)[r]{\strut{}$0.4$}}%
      \put(814,1823){\makebox(0,0)[r]{\strut{}$0.6$}}%
      \put(814,2196){\makebox(0,0)[r]{\strut{}$0.8$}}%
      \put(814,2569){\makebox(0,0)[r]{\strut{}$1$}}%
      \put(946,484){\makebox(0,0){\strut{}$0$}}%
      \put(1925,484){\makebox(0,0){\strut{}$0.2$}}%
      \put(2903,484){\makebox(0,0){\strut{}$0.4$}}%
      \put(3882,484){\makebox(0,0){\strut{}$0.6$}}%
      \put(4860,484){\makebox(0,0){\strut{}$0.8$}}%
      \put(5839,484){\makebox(0,0){\strut{}$1$}}%
    }%
    \gplgaddtomacro\gplfronttext{%
      \csname LTb\endcsname%
      \put(176,1636){\rotatebox{-270}{\makebox(0,0){\strut{}$\rho_{\text{hybrid loop}}$}}}%
      \put(3392,154){\makebox(0,0){\strut{}$\omega [a_{\text{s}}^{-1}]$}}%
      \csname LTb\endcsname%
      \put(4852,2396){\makebox(0,0)[r]{\strut{}Free}}%
      \csname LTb\endcsname%
      \put(4852,2176){\makebox(0,0)[r]{\strut{}$R=\SI{1}{a_{\text{s}}}$}}%
      \csname LTb\endcsname%
      \put(4852,1956){\makebox(0,0)[r]{\strut{}$R=\SI{2}{a_{\text{s}}}$}}%
      \csname LTb\endcsname%
      \put(4852,1736){\makebox(0,0)[r]{\strut{}$R=\SI{3}{a_{\text{s}}}$}}%
      \csname LTb\endcsname%
      \put(4852,1516){\makebox(0,0)[r]{\strut{}$R=\SI{4}{a_{\text{s}}}$}}%
      \csname LTb\endcsname%
      \put(4852,1296){\makebox(0,0)[r]{\strut{}$R=\SI{5}{a_{\text{s}}}$}}%
    }%
    \gplbacktext
    \put(0,0){\includegraphics{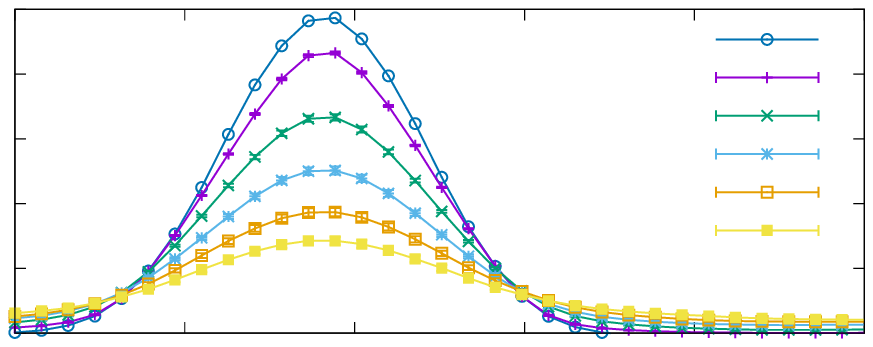}}%
    \gplfronttext
  \end{picture}%
\endgroup